\documentclass[12pt]{article}

\def\bb{\begin{equation}}
\def\ee{\end{equation}}
\def\pt{\partial}

\def\const{\hbox{const}}

\def\ve{\varepsilon}

\begin{document}

\begin{center}{\Large\bf Scattering of solitons on resonance}\footnote{This work was supported by RFBR 03-01-00716, grant for Sci. Schools 1446.2003.1 and INTAS 03-51-4286}
\\
\vspace{3mm}
{O M Kiselev, Institute of Mathematics of Ufa \\ 
Sci. Centre of RAS, E-mail: ok@ufanet.ru
\\
S G Glebov, Ufa State Petroleum Technical University, \\
E-mail: glebskie@rusoil.net}
\end{center}
\centerline{September 17, 2003}

\begin{abstract}
We investigate a propagation of solitons for
nonlinear Schrodinger equation under small driving force. The
driving force passes the resonance. The process of scattering  on
the resonance leads to changing of number of solitons. After the
resonance the number of solitons  depends on the amplitude of the
driving force.
\end{abstract}

Nonlinear Schrodinger equation (NLSE) is a mathematical model for
wide class of wave phenomenons from signal propagation into optical
fibre \cite{Kelley, Talanov} to surface wave propagation
\cite{Zakharov}. This equation is integrable by inverse scattering
transform method \cite{TeoriyaSolitonov} and can be considered as an
ideal model equation. The perturbations of this ideal model lead to
nonintegrable equations. Here we consider such nonintegrable example
which is NLSE  perturbed driving force.
\par
The most known class of the solutions of NLSE is solitons
\cite{TeoriyaSolitonov}. The structure of this kind of solutions is
not changed in a case of nonperturbed NLSE. The perturbations
usually  lead to modulation of parameters of solitons \cite{Kaup,
Karpman-Maslov}. Number of solitons does not change.
\par
In this work we investigate a new effect called scattering of
solitons on resonance. We consider the process of scattering in
detail and obtain the connection formula between pre-resonance and
post-resonance solutions. In general case the passage through
resonance leads to changing of the number of solitons. This effect
is based on the soliton generation due to passage through resonance
by external driving force \cite{Glebov-Kiselev-Lazarev}.
\par
We found that the scattering of solitary waves on resonance 
is a general effect for nonlinear equations described the wave propagation. 
In this work we investigate this effect for the simplest model. 
It allows to show the essence of this effect without unnecessary details.
\par
This paper has the following structure. The first section contains
the statement of the problem and the main result. The second section
contains the asymptotic construction in the pre-resonance domain. In
the third section we construct  the asymptotic solution in the
neighborhood of the resonance curve. The fourth section of the paper
is devoted to construction of the post-resonance asymptotics.
Asymptotics are constructed by multiple scale method \cite{JK} and
matched \cite{Il'in}.
\par

\section{Statement of the problem and  result}

Let us consider the perturbed NLSE
\bb
i \pt_t \Psi + \pt_{x}^2 \Psi + |\Psi|^2 \Psi = \ve^2 f
e^{iS/\ve^2},\qquad 0< \ve \ll 1. \label{sh}
\ee
The phase of the driving force is $S/\ve^2 = \omega t$. The
amplitude $f=f(\ve x)$ is a smooth and rapidly vanished function.
\par
In the simplest case the phase is linear function with respect to $t$
$S/\ve^2=\omega t,\ \omega=\const$. In general situation the
constant frequency of the driving force does not lead to scattering
of solitons. Let us investigate the driving force with slowly
varying frequency. The most simplest dependence on $t$ for $\omega$
has a form $\omega=\ve^2 t/2$. The amplitude $f$ of the driving
force admit an additional dependency on $\ve^2 t$ but it leads to
complicated formulas and no more.
\par
Let us formulate the result of this  work. Below we use the
following variables $x_j=\ve^j x,\ t_j=\ve^j t,\ j=1,2$.
\par
Let the asymptotic solution of (\ref{sh}) be
$$
\Psi(x,t,\ve)=\ve \stackrel{1}{u}(x_1,t_2)+O(\ve^2)\quad
\mbox{as}\quad t_2 < 0,
$$
where  $\stackrel{1}{u}(x_1,t_2)$ satisfies
$$
\pt_{t_2}\stackrel{1}{u}+\pt_{x_1}^2 \stackrel{1}{u}+|
\stackrel{1}{u}|^2 \stackrel{1}{u}=0
$$
and initial condition
$$
\stackrel{1}{u}|_{t_2=t^0}=h_1(x_1),\quad t^0=\const < 0.
$$
Then in the domain $t_2>0$ the asymptotic solution of (\ref{sh}) has
a form
\begin{eqnarray}
\Psi(x,t,\ve)=\ve \stackrel{1}{v}(x_1,t_2)+O(\ve^2),
\end{eqnarray}
where $\stackrel{1}{v}(x_1,t_2)$ is a solution of NLSE with initial
condition
$$
\stackrel{1}{v}|_{t_2=0}=\stackrel{1}{u}(x_1,0)+(1-i)\sqrt{\pi}f(x_1).
$$
Let us explain the result for soliton solution. If in the domain
$t_2<0$ the solution has   $N$-soliton form then in the domain
$t_2>0$ the number of solitons is defined by initial condition for
$\stackrel{1}{v}$.

\section{Incident wave}
\par
In this section we construct the asymptotic solution of equation
(\ref{sh}) in pre-resonance domain. This solution contains two
parts. The first part is a specific solution of the nonhomogeneous
equation. This solution oscillates with the frequency of the driving
force. The amplitude is determined by an algebraic equation. The
second part of the solution is a solution of the homogeneous
equation. The solution contains undefined function due to
integration. This undefined function usually determines by initial
condition for Cauchy problem.
\par
We construct the formal asymptotic solution of the form
\begin{eqnarray}
\Psi(x,t,\ve)=\ve \stackrel{1}{u}(x_1,t_2)+\ve^3\stackrel{3}{u}(x_1,t_2)+
\ve^{2}\stackrel{2}{B}(x_1,t_2)\exp(iS/\ve^2)+ \nonumber\\
\ve^4\bigg(\stackrel{4}{B}_{1}(x_1,t_2)\exp(iS/\ve^2)+
\stackrel{4}{B}_{-1}(x_1,t_2)\exp(-iS/\ve^2)\bigg)
\label{external-anzats-2}\\
+\ve^{5}\stackrel{5}{B}_2(x_1,t_2)\exp(2iS/\ve^2).\nonumber
\end{eqnarray}
\par
To determine the coefficients of the asymptotics substitute
(\ref{external-anzats-2}) into equation (\ref{sh}). It yields
\begin{eqnarray*}
\ve^2\bigg(-S'\stackrel{2}{B}-f\bigg)\exp(iS/\ve^2)+
\ve^3\bigg(i\stackrel{1}{u}_{t_2}+\stackrel{1}{u}_{x_1x_1}+
|\stackrel{1}{u}|^2\stackrel{1}{u}\bigg)
\\
+\ve^{4}\bigg(\bigg(-S'\stackrel{4}{B}_{1}+
i\stackrel{2}{B}_{t_2}+
\stackrel{2}{B}_{x_1x_1}+
2|\stackrel{1}{u}|{}^2\stackrel{2}{B}\bigg)\exp(iS/\ve^2)+\\
+\bigg(S'\stackrel{4}{B}_{-1}+
\stackrel{1}{u}{}^2\stackrel{2}{B}{}^*\bigg)\exp(-iS/\ve^2)\bigg)
\\
\ve^5\bigg(
i\stackrel{3}{u}_{t_2}+\stackrel{3}{u}_{x_1x_1}
+2|\stackrel{1}{u}|^2\stackrel{3}{u}+
\stackrel{1}{u}{}^2\stackrel{3}{u}{}^*+
2\stackrel{1}{u}|\stackrel{2}{B}|{}^2+\nonumber\\
\bigg(-2S'\stackrel{5}{B}_{2}+
\stackrel{1}{u}{}^*\stackrel{2}{B}{}^2\bigg)\exp(2iS/\ve^2)\bigg)
= {\cal O}(\ve^6R(t_2,x_1)).
\end{eqnarray*}
\par
The residue part of the asymptotics  has a form
$$
R(t_2,x_1)=O(|\stackrel{2}{B}|^3+\ve^3|\stackrel{3}{u}|^3+\ve^6|\stackrel{4}{B}_{1}|^3+\ve^6|\stackrel{4}{B}_{-1}|^3+\ve^9|\stackrel{5}{B}_2|^3).
$$
\par
Collect the terms with the same order of $\ve$ up to  the order of
$\ve^5$ and reduce similar  terms. It yields  differential equations
for $\stackrel{1}{u}, \stackrel{3}{u}$ and algebraic equations for
$\stackrel{2}{B}$, $\stackrel{4}{B}_{\pm1}$ and
$\stackrel{5}{B}_{2}$.
\begin{eqnarray}
i\stackrel{1}{u}_{t_2}+\stackrel{1}{u}_{x_1x_1}+
|\stackrel{1}{u}|^2\stackrel{1}{u}=0,\nonumber\\
i\stackrel{3}{u}_{t_2}+\stackrel{3}{u}_{x_1x_1}+
2|\stackrel{1}{u}|^2\stackrel{3}{u}+
\stackrel{1}{u}{}^2\stackrel{3}{u}{}^*=
-2|\stackrel{2}{B}|^2\stackrel{1}{u},\nonumber
\end{eqnarray}
\begin{eqnarray}
-S'\stackrel{2}{B}=f,\nonumber \\
-S'\stackrel{4}{B}_1=i\stackrel{2}{B}_{t_2}+\stackrel{2}{B}_{x_1x_1}+
|\stackrel{1}{u}|^2\stackrel{2}{B}, \nonumber\\
S'\stackrel{4}{B}_{-1}=-\stackrel{1}{u}{}^2 \stackrel{2}{B}{}^*,\nonumber\\
-2S'\stackrel{5}{B}_2=\stackrel{1}{u}{}^*\stackrel{2}{B}{}^2.\nonumber
\end{eqnarray}
\par
The coefficients $\stackrel{1}{u},  \stackrel{3}{u}$  are uniquely
determined by initial conditions at the moment $t_2=t^0$. We suppose
that $t^0=\const<0$ and
$$
\stackrel{1}{u}|_{t_2=t^0} = h_1(x_1);\quad
\stackrel{3}{u}|_{t_2=t^0} = h_3(x_1);
$$
where functions $h_1, h_3$ are smooth and rapidly vanish as $|x_1|\to
\pm\infty$.
\par
The coefficients of the representation (\ref{external-anzats-2})
have a singularity as $S'\to 0$. The order of singularity of
$\stackrel{j}{B}_{k}$ is easy calculated.
$$
\stackrel{2}{B}=O(t^{-1}),\quad \stackrel{4}{B}_1=O(t^{-3}).
$$
\par
To determine the asymptotics of $\stackrel{3}{u}$ as $t_2\to-0$
we construct the solution of the form
$$
\stackrel{3}{u}=t_2^{-1}\stackrel{3}{u}{}^{(-1,0)}(x_1,t_2)+
\ln|t_2|\stackrel{3}{u}{}^{(0,1)}(x_1,t_2)+
t_2\ln|t_2|\stackrel{3}{u}{}^{(1,1)}(x_1,t_2)+
\widehat{\stackrel{3}{u}}(x_1,t_2).
$$
Substitute this representation into equation for $\stackrel{3}{u}$
and collect the terms of the same order with respect to $t_2$. It
yields equations for coefficients of the asymptotics
\begin{eqnarray}
\stackrel{3}{u}{}^{(-1,0)}=i2|f|^2\stackrel{1}{u},\nonumber\\
\stackrel{3}{u}{}^{(0,1)}=-iL(\stackrel{3}{u}{}^{(-1,0)}),\nonumber\\
\stackrel{3}{u}{}^{(1,1)}=-iL(\stackrel{3}{u}{}^{(0,1)}),\nonumber\\
L(\widehat{\stackrel{3}{u}})=it_2\ln|t_2|L(\stackrel{3}{u}{}^{(1,1)})+
i\stackrel{3}{u}{}^{(1,1)}.\label{hat-u3-eq}
\end{eqnarray}
Here  $L(u)$ is a linear operator of the form
$$
L(u)=i\pt_{t_2}u+\pt_{x_1}^2 u+2|\stackrel{1}{u}|^2 u +\stackrel{1}{u}{}^2u^*.
$$
Functions $\stackrel{3}{u}{}^{(-1,0)}$,
$\stackrel{3}{u}{}^{(0,1)}$ and $\stackrel{3}{u}{}^{(1,1)}$ are
determined from algebraic equations. These functions are bounded
as $-\const<t_2\le0,\  \const>0$.
\par
The function $\widehat{\stackrel{3}{u}}$ is a solution of
nonhomogeneous  linearized Schrodinger equation. The right hand side
of the equation is a smooth function as $-\const<t_2\le0,\
\const>0$. The solution of this equation can be obtained using
results of \cite{Keen-McL}. In particularly if $\stackrel{1}{u}$ is
N-solitons solution of NLSE then exists the bounded solution of
nonhomogeneous  linearized Schrodinger equation (\ref{hat-u3-eq}) as
$-\const<t_2\le0,\  \const>0$.
\par
Coefficients of (\ref{external-anzats-2}) have  singularity at
$t_2=0$. After substitution  (\ref{external-anzats-2}) into equation
(\ref{sh}) we obtain a residue part. This residue part increases as
$t_2 \to 0$. The domain of validity of (\ref{external-anzats-2}) is
determined by
$$
\ve^6R(t_2,x_1)=o(\ve),\quad \ve\to0.
$$
It yields
$$
-t_2\gg\ve \quad \hbox{or} \quad -t \gg \ve^{-1}.
$$
\par

\section{Scattering}

In the neighborhood of the point $t_2=0$ the frequency of the
driving force becomes resonant. It leads to changing of behaviour of
the system described by (\ref{sh}). The forced mode of oscillations
changes by a resonant mode. Formally it means representation
(\ref{external-anzats-2}) is not valid.
\par
In this part of the work we construct another representation for the
solution of equation (\ref{sh}). This representation is valid in the
neighborhood of the resonance line $t_2=0$.
\begin{eqnarray}
\Psi(x,t,\ve)=\ve\stackrel{1}{w}(x_1,t_1)\big)+
\ve^{2}\stackrel{2}{w}(x_1,t_1)+\nonumber\\
\ve^3\ln\ve\stackrel{3,1}{w}(x_1,t_1)+
\ve^{3}\stackrel{3}{w}(x_1,t_1) \qquad
 \ve \to 0. \label{internal-anzats}
\end{eqnarray}
Here we use a new scaled variable $t_1=t_2/\ve$. Representation
(\ref{internal-anzats}) is matched with (\ref{external-anzats-2}).
It means these formulas are equivalent up to value $o(\ve^5)$ as
$t_2\to-0$. The coefficients $\stackrel{m}{w}$ of
(\ref{internal-anzats}) are determined by ordinary differential
equations (\ref{inw0}), (\ref{inw1}), (\ref{inw2}) and matching
conditions.
\par
To obtain the behaviour of the coefficients of
(\ref{internal-anzats}) as $t_1\to-\infty$ match
(\ref{internal-anzats}) with (\ref{external-anzats-2}). Write
(\ref{external-anzats-2}) in terms of $t_1$
$$
\Psi(x,t,\ve)=\ve\bigg(\stackrel{1}{u}(x_1,0)-\big(t_1^{-1}f+
it_1^{-3}f\big)\exp(iS/\ve^2)\bigg)
+
$$
$$
\ve^2\bigg(\pt_{t_2}\stackrel{1}{u}(x_1,t_2)|_{t_2=0} t_1+
t_1^{-1}i|f|^2\stackrel{1}{u}(x_1,0)+O\big(t_1^{-2}\big)\bigg)+
$$
$$
\ve^3\ln\ve\bigg(-iL(2i|f|^2\stackrel{1}{u})|_{t_2=0}+o(1)\bigg)+
$$
$$
 +\ve^3\bigg({1\over2}\pt_{t_2}^2\stackrel{1}{u}(x_1,t_2)|_{t_2=0}
t_1^2+\widehat{\stackrel{3}{u}}(x_1,0)+o(1)\bigg), \quad
1\ll-t_1\ll\ve^{-1},\,\,\ve\to0.
$$
\par
To obtain equations for coefficients of (\ref{internal-anzats})
substitute (\ref{internal-anzats}) into equation (\ref{sh}). It
yields
\begin{eqnarray*}
\ve^2\bigg((\pt_{t_1}\stackrel{1}{w}-f\exp(iS/\ve^2)\bigg)+
\ve^3\bigg(\pt_{t_1}\stackrel{2}{w}+ \pt_{x_1}^2\stackrel{1}{w}+
\gamma|\stackrel{1}{w}|\stackrel{1}{w}\bigg) +\\
\ve^4\bigg(\pt_{t_1}\stackrel{3}{w}+ \pt_{x_1}^2\stackrel{2}{w}+
+\stackrel{1}{w}{}^2\stackrel{2}{w}{}^* +
2\gamma|\stackrel{1}{w}|\stackrel{2}{w}\bigg)=O(\ve^5\rho(t_1,x_1,\ve)).
\end{eqnarray*}
The function $\rho(t_1,x_1,\ve)$ can be represented in the form
$$
\rho(t_1,x_1,\ve)=O(|\stackrel{1}{w}|^2\stackrel{3}{w}+\pt_{x_1}^2\stackrel{3}{w}
+\ve|\stackrel{2}{w}|^3+\ve^4|\stackrel{3}{w}|^3).
$$
Collect the terms of the same order with respect to $\ve$. As
result we obtain the equation for $\stackrel{1}{w}$
\begin{eqnarray}
i\pt_{t_1}\stackrel{1}{w}=f\exp(it_1^2/2). \label{inw0}
\end{eqnarray}
The matching conditions give
$\stackrel{1}{w}=\stackrel{1}{u}(x_1,0)\,\,\, t_1\to-\infty$. The
solution of this problem is represented in terms of Fresnel
integral
\bb
\stackrel{1}{w}=\stackrel{1}{u}(x_1,0) -if(x_1)\int_{-\infty}^{t_1}\exp(i\theta^2/2)d\theta.
\label{leading-order-of internal-exp}
\ee
Equations for higher-order terms are
\bb
i\pt_{t_1}\stackrel{2}{w}=-\pt_{x_1}^2\stackrel{1}{w}+
|\stackrel{1}{w}|^2\stackrel{1}{w}, \label{inw1}
\ee
\bb
i\pt_{t_1}\stackrel{3,1}{w}=0, \label{inw21}
\ee
\bb
i\pt_{t_1}\stackrel{3}{w}=-\pt_{x_1}^2\stackrel{2}{w}-
2|\stackrel{1}{w}|^2\stackrel{2}{w}-
\stackrel{1}{w}{}^2\stackrel{2}{w}{}^*. \label{inw2}
\ee
\par
The higher-order terms satisfy fist order ordinary differential
equations with respect to $t_1$. The spatial variable $x_1$ is a
parameter in these equations. The solutions of these equation are
uniquely defined by terms of the order of $1$ in asymptotics as
$t_1\to-\infty$. The asymptotics as $t_1\to-\infty$ is obtained
by matching
$$
\stackrel{2}{w}=\pt_{t_2}\stackrel{1}{u}(x_1,t_2)|_{t_2=0} t_1+o(1);
$$
$$
\stackrel{3}{w}{}^1=-iL(2i|f|^2\stackrel{1}{u})|_{t_2=0}+o(1),
$$
$$
\stackrel{3}{w}={1\over2}\pt_{t_2}^2\stackrel{1}{u}(x_1,t_2)|_{t_2=0} t_1^2+\widehat{\stackrel{3}{u}}(x_1,0)+o(1).
$$
To determine the behaviour of the solution after resonance we need
to calculate the asymptotics as $\tau \to +\infty$ of the
coefficients for representation (\ref{internal-anzats}).
Calculations give
$$
\stackrel{1}{w}(x_1,t_1)=\stackrel{1}{u}(x_1,0)-i f(x_1)\bigg[ic_1 +
{\exp(it_1^2/2)\over it_1}+O(t_1^{-3})\bigg],
$$
where $c_1 = (1-i)\sqrt{\pi}$.
\par
Denote by
$$
\stackrel{1}{w}(x_1,t_1)|_{t_1\to\infty}=\stackrel{1}{w}_0(x_1).
$$
The function $\stackrel{2}{w}(x_1,t_1)$ has the asymptotics of the
form
$$
\stackrel{2}{w}(x_1,t_1)=t_1\stackrel{2}{w}{}_{1}(x_1) + \stackrel{2}{w}{}_0(x_1)+ g_1(x_1){\exp(it_1^2/2)\over
it_1^2} +O(t_1^{-4}),
$$
where
$$
\stackrel{2}{w}{}_1=-\pt_{x_1}^2\stackrel{1}{w}_0+|\stackrel{1}{w}_0|^2\stackrel{1}{w};
$$
$$
\stackrel{2}{w}{}_{0}(x_1)=\lim_{t_1\to\infty}\bigg(
\int_{-\infty}^{t_1} \big[\pt_{x_1}^2\stackrel{0}{w}(x_1,\theta)+
|\stackrel{0}{w}(x_1,\theta)|^2\stackrel{0}{w}(x_1,\theta)\big]d\theta
-\stackrel{2}{w}_{1}t_1\bigg),
$$
$g_1(x_1)=k_1 \pt_{x_1}f+k_2|f|^2 f$, $k_1$ and $k_2$ are
constants.
$$
\stackrel{3}{w}(x_1,t_1)=t_1^2\stackrel{3}{w}{}_{2}(x_1)
+ o(t_1^2).
$$
where
$$
\stackrel{3}{w}{}_2(x_1)=i\left(\pt_{x_1}^2\stackrel{2}{w}_1
+2|\stackrel{1}{w}_0|^2\stackrel{2}{w}_1 +
\stackrel{1}{w}{}_0^2\stackrel{2}{w}{}_1^*\right),
$$
\par
Representation (\ref{internal-anzats}) of the solution for
(\ref{sh}) is valid as
$$
\ve^5\rho(t_1,x_1,\ve)=o(\ve).
$$
The determined above behaviour of coefficients of asymptotics
(\ref{internal-anzats}) give the domain of validity with respect to
$t_1$
$$
|t_1| \ll \ve^{-1} \quad \hbox{or} \quad |t| \ll \ve^{-2}.
$$

\section{Scattered wave}

In this section we construct the asymptotic solution of equation
(\ref{sh}) after the resonance.  The leading-order term of the
solution satisfies NLSE and depends on $x_1,t_2$ as well as before
resonance. But this leading-order term  is determined by another
solution of NLSE which contains generally speaking another number of
solitons. This number depends on a condition on the resonance curve
$t_2=0$.
\par
After resonance we construct the asymptotic solution of the form
\begin{eqnarray}
\Psi(x,t,\ve)=\ve\stackrel{1}{v}(x_1,t_2)+
\ve^2\stackrel{2}{v}(x_1,t_2)+
\nonumber\\
\ve^2\stackrel{2}{A}(t_2,x_1)\exp(iS/\ve^2)+\ve^4(\stackrel{4}{A}_1(t_2,x_1)\exp(iS/\ve^2)+
\nonumber\\
\stackrel{4}{A}_{-1}(t_2,x_1)\exp(-iS/\ve^2)).
\label{external-anzats2}
\end{eqnarray}
\par
Substitute this representation into  (\ref{sh}):
\begin{eqnarray*}
\ve^2(-S'\stackrel{2}{A}-f)\exp(iS/\ve^2) +\ve^3(\pt_{t_2}\stackrel{1}{v}+\pt_{x_1}^2\stackrel{1}{v}+ |\stackrel{1}{v}|^2\stackrel{1}{v})+
\\
\ve^4(\pt_{t_2}\stackrel{2}{v}+\pt_{x_1}^2\stackrel{2}{v}+2 |\stackrel{1}{v}|^2\stackrel{2}{v}+ \stackrel{1}{v}^2\stackrel{2}{v}{}^*+
\\ (-S'\stackrel{4}{A}_{1}+\pt_{t_2}\stackrel{2}{A}+ \pt_{x_1}^2\stackrel{2}{A}+2|\stackrel{1}{v}|^2\stackrel{2}{A})\exp(iS/\ve^2)+\\ (S'\stackrel{4}{A}_{-1}+\stackrel{1}{v}{}^2\stackrel{2}{A}{}^*)\exp(-iS/\ve^2))=\ve^5 r(t_2,x_1,\ve).
\end{eqnarray*}
Here $r(t_2,x_1,\ve)$ depends on coefficients of the asymptotics
(\ref{external-anzats2}). This dependence is easy calculated. The
coefficients $\stackrel{2}{A}$, $\stackrel{4}{A}_1$ è
$\stackrel{4}{A}_{-1}$ have singularity on the resonance curve. To
determine the domain of validity of (\ref{external-anzats2}) we need
to derive the explicit formula for $r$
$$
r(t_2,x_1,\ve)=O(1+\ve|\stackrel{2}{A}|^3+\ve\ln|\ve|+\ve^7(|\stackrel{4}{A}_1|^3
+|\stackrel{4}{A}_{-1}|^3)).
$$
\par
Collect the terms of the same order of small parameter and the same
exponents. It yields the equations for coefficients of representation
(\ref{external-anzats2}).
\bb
\pt_{t_2}\stackrel{1}{v}+\pt_{x_1}^2\stackrel{1}{v}+ |\stackrel{1}{v}|^2\stackrel{1}{v}=0;
\label{sh-2}
\ee
\bb
\pt_{t_2}\stackrel{2}{v}+\pt_{x_1}^2\stackrel{2}{v}+2 |\stackrel{1}{v}|^2\stackrel{2}{v}+ \stackrel{1}{v}^2\stackrel{2}{v}{}^*=0
\label{lsh-1}
\ee
$$
-S'\stackrel{2}{A}=f;
$$
$$
-S'\stackrel{4}{A}_{1}=-\pt_{t_2}\stackrel{2}{A}- \pt_{x_1}^2\stackrel{2}{A}-2|\stackrel{1}{v}|^2\stackrel{2}{A};
$$
$$
S'\stackrel{4}{A}_{-1}=-\stackrel{1}{v}{}^2\stackrel{2}{A}{}^*.
$$
\par
Initial conditions for differential equations for $\stackrel{1}{v}$
are obtained by matching. These conditions are evaluated on the
resonance curve $t_0=0$.
\bb
\stackrel{1}{v}|_{t_2=0}=\stackrel{1}{u}(x_1,0)+(1-i)\sqrt{\pi}f(x_1);
\label{ic1}
\ee
\bb
\stackrel{2}{v}|_{t_2=0}=\stackrel{2}{w}_{0}(x_1).
\label{ic2}
\ee
\par
The residue part $\ve^5r(t_2,x_1,\ve)=o(\ve)$ as $t_2\gg\ve$. This
condition is determined the domain of validity for
(\ref{external-anzats2}).
\par
Formula (\ref{ic1}) is connection formula for solution before and
after the resonance. Additional term $(1-i)\sqrt{\pi}f(x_1)$ leads
to changing of the solution after passage through the resonance.
\par
{\bf Acknowledgments.} We are grateful to  I.V. Barashenkov, L.A.
Kalyakin and B.I.Suleimanov for helpful comments.

\end{document}